\documentstyle[12pt]{article}

\newcommand{\ket}[1]{\mbox{$\vert {#1} \rangle $}}
\newcommand{\bra}[1]{\mbox{$\langle {#1} \vert $}}

\begin{document}
\vspace*{-1.5cm}
 
May 1997 \hfill   ULB--TH 10/97 \\
\phantom{z}\hfill hep-th/9705115
 
\vskip 3cm
\centerline{\bf ON THE BLACK HOLE UNITARITY ISSUE\footnote{Talk presented at
the  Workshop on Frontiers in Field theory, Quantum Gravity and String
Theory (December 1996) Puri, India.}}
\vskip1cm 
\begin{center}
Fran\c{c}ois~Englert\footnote{ e-mail: fenglert@ulb.ac.be}\\[.3cm] {\it
Service de Physique Th\'eorique\\ Universit\'e Libre de Bruxelles, Campus
Plaine, C.P. 225\\ Boulevard du Triomphe, B-1050 Bruxelles, Belgium }
\vskip .5cm
\end{center}
\vskip 1cm 

\begin{abstract}

\noindent  
I discuss features required for preserving unitarity in black hole
decay and concepts underlying such a perspective. Unitarity requires that
correlations  extend on the scale of the   horizon.  I show, in a toy model 
inspired by string theories, that such    correlations can indeed 
arise. The model suggests that, after  a time of order $4M \ln M$ following  the
onset of Hawking radiation, quantum effects could  maintain  throughout the 
decay a collapsing star within a   Planck distance of its Schwarzschild radius. 
In this way information loss would be avoided. The concept of  black hole 
``complementarity'',  which could reconcile these macroscopic departures from
classical physics with the equivalence principle, is interpreted in terms of
weak values of quantum operators.

\end{abstract} 

\newpage
 
\section{Introduction}

In view of the successful computation of the quantum degeneracy of some extreme
black holes in string theory \cite{vafa,callan}, it may seem likely that
further developments   will   provide the solution to the problem
posed by black hole decay and that there is not much   to gain from
general considerations. However, even if  string theory would
ultimately  lead to a black hole decay   consistent with unitarity, the
nature of such a decay  is presently unclear. It may
therefore be of some help to investigate   features required for
preserving unitarity in   black hole decay and    concepts underlying such a
perspective.

I shall first briefly review the unitarity problem and stress the highly non
local correlations which would be needed to avoid  information loss. I
shall then examine in a toy model how such non local effect can ``materialise''
the black hole horizon and how such a materialisation  suggests
that the degrees of freedom of a collapsing star are recovered in
the Hawking radiation.  I shall then discuss the black hole complementarity issue
and its  possible quantum significance. 

\section{The Unitarity Issue}

 The Bekenstein assumption \cite{bek} that   a black hole of mass $M$ has an
entropy proportional to the  area $A=4\pi M^2$ of its event horizon, combined
with the Hawking computation of the black hole  radiation temperature
\cite{haw} $T_H = 1/8\pi M$, yields for this entropy $S$ the value
\begin{equation}
\label{entropy}
  S= A/4 + C  
\end{equation}
where $C$ is an unknown integration constant. This is an immediate consequence
of the identity
\begin{equation}
\label{mass} 
 dM=T_H dS. 
\end{equation}

The original derivation of the Hawking radiation \cite{haw} was based on
conventional local field theory and did not take into account the gravitational
backreaction. According to these assumptions, the emitted quanta are  only  
correlated  to states formed out of degrees of freedom in the horizons
vicinity and     no information about the infalling matter is stored in
the radiation itself.   Complete evaporation of the black hole would  then
inevitably lead in the semi-classical approximation to a density matrix out of
any initial state, even out of a pure quantum state, and thus to a violation       of
unitarity \cite{haw2}. An   alternative is however conceivable if, when the
black hole reaches the Planck regime where semi-classical considerations become
unreliable, evaporation would stop and leave a remnant with very high and
possibly infinite degeneracy to which the radiation would remain
correlated \cite{remnant}. If a statistical interpretation of the black hole
entropy were available, infinite degeneracy would mean an infinite integration
constant in  Eq.(\ref{entropy}).

It was first suggested by 't Hooft that the semi-classical approximation breaks
down already at macroscopic scales and that unitarity could be maintained without
the remnant hypothesis\cite{thooft}. Information must then be
transmitted to the radiation. Causality implies that   information about the
collapsing star should not be trapped inside  the horizon. This  means
that macroscopic correlations   must  be present to turn the horizon into a
system where information can be deposited. Such
``materialisation'' of the horizon  can only appear at the quantum level and has
dramatic implications.  Firstly, the black hole entropy is interpretable as a
counting of a finite  number of   states, and thus   the integration 
constant in the area entropy (\ref{entropy}) must be   zero or finite.
Furthermore, any system defined in a finite volume has an energy   limited by
black hole formation and the number of states in this volume is   bounded by the
logarithm of the area entropy of the surrounding surface. This is at odds 
with   local field theory which would lead, even in presence of a Planckian
cut-off, to an entropy  proportional to the volume and hence to an infinite
constant of integration in the thermodynamic limit. Thus a theory underlying
unitary black hole decay is expected to be genuinely non local. Finally, the  
materialisation of a horizon, where   the classical curvature is vanishingly
small in the large $M$ limit,   violates the equivalence principle. This  
motivated the introduction of a complementarity hypothesis according to which  
an inertial observer who would cross the horizon would not feel its material
structure \cite{suss}.
 
\section{String Instantons and Non Local Effects.}

\subsection{Entropy and Temperature.}

I shall show that, in a toy model inspired from string theory, non local effect
can modify both the black hole temperature and the value of  its entropy 
(\ref{entropy}). These   changes affect drastically the properties of quantized
matter in the black hole background. The model   discussed here is not meant to
be a realistic one but its interest   is to reveal a mechanism by which
unitarity could be maintained in black hole evaporation.

The possibility of departing from $A/4$   can be understood in
thermodynamic terms. The differential mass formula \cite{bch} which
generalises  (\ref{mass}) for black holes surrounded  by static matter
can be written as \cite{ce}   
\begin{equation} 
 \label{dmf} 
\delta M_{tot}={\kappa\over 2\pi}{\delta A\over 4} + 
\sum_i\partial_{\lambda_i}H_{matter}\delta\lambda_i 
\end{equation}
 where $M_{tot}$ is the total mass, $\kappa$   the surface gravity of the hole,
$A$ the area of the event horizon and the $\lambda_i$ are all the parameters in
the matter action. This identity is derived from classical physics and
cannot involve  the Planck constant. The Bekenstein assumption  that the black
hole contributes to the entropy amounts to interpret Eq.(\ref{dmf}) as the
expression of the first principle of thermodynamics:  $A/4$ is  the entropy up
to a multiplicative constant proportional to $\hbar^{-1}$ and the temperature
must then be proportional to $\hbar$. The $\partial_{\lambda_i}H_{matter}$ are
generalised forces.

It is well known that to compute thermal correlation functions and
partition functions in field theory in flat Minkowski space-time one can
use path integrals in periodic imaginary time.  The period $\beta$ is the
inverse temperature and can be chosen freely. This method was generalised  to
compute matter correlation functions in static curved backgrounds.  For the
Schwarzschild black hole, possibly surrounded by matter, the analytic
continuation to imaginary time defines a Euclidean background everywhere except
at the analytic continuation of the horizon, namely the 2-sphere at $r=2M$.  
Gibbons and Hawking \cite{gh1} extended the analytic continuation to the
gravitational action, restricting the hitherto ill-defined path integral over
metrics to a saddle point in the Euclidean section.   To constitute such a
saddle the Euclidean black hole must be regular given that a singularity at
$r=2M$ would invalidate the solution of the Euclidean Einstein equations. This
implies a unique   Hawking temperature $T_H$ which, in natural units, is always
equal to $\kappa / (2\pi)$.   Thus one recovers from Eq.(\ref {dmf})  
the  area entropy  (\ref{entropy}).    
 
This entropy  is not affected by mass surrounding
the black hole and would therefore  seem   to depend only on
the black hole mass. But this need not be the case: a different relation between
entropy and area arises when a conical singularity is generated in the
Euclidean section at $r=2M$ \cite{ehw1}.
 
As pointed out by many authors,    a conical singularity at
$r=2M$ modifies the Euclidean periodicity of the black hole and hence its
temperature. If  this singularity would arise from a source term in the
Euclidean Einstein equations,  Eq.(\ref {dmf})  would remain valid and could be
written as  
\begin{equation} 
\label{dmfnew}
\delta M_{tot}=T\delta [(1-\eta) {A\over 4}] + T{A\over 4}\delta \eta +
\sum_i\partial_{\lambda_i}H_{matter}\delta\lambda_i,
\end{equation} 
where $\eta$ is the deficit angle and the temperature $T$ is related to the
Hawking value $T_H$ by $T=T_H (1-\eta)^{-1}$. It follows from  Eq.(\ref{dmfnew})
that a new generalised force $X_{\eta}$, conjugate to $\eta$,  
\begin{equation} 
\label{force}
X_{\eta}=T {A\over 4}
\end{equation}
must appear and that the entropy of the hole would become
\begin{equation} 
\label{areanew}
 S =(1-\eta) {A\over 4},   
\end{equation}
independent  of the   surrounding matter. I now show that a deficit angle can be
generated by a string instanton and that Eqs.(\ref {force}) and (\ref
{areanew}) obtain with  
 $\eta$ determined by the string tension. 

For simplicity I shall describe here only the case of a pure black hole of mass
$M$. The general case is treated elsewhere \cite{ehw2}. In presence of a
Nambu-Goto string the Euclidean action is
\begin{eqnarray}
\nonumber
  I &=&-{1\over 16\pi}\int_{  M}\ \sqrt{ g }
R +{1\over 8\pi}\int_{\partial   M}\ \sqrt{ h }K \\
\label{action}
&&- {1\over
8\pi}\int_{(\partial   M)_\infty}\ \sqrt{ h_0 } K_0 + \mu \int\
d^2\sigma \sqrt{\gamma}.
\end{eqnarray} 
Here $\mu$ is the string tension and $\gamma$ the  determinant   of the induced
metric on the world sheet. The latter is taken to have the topology of a
2-sphere. The variation of this action with respect to the metric
  gives the   Einstein equations and the variations with respect to the string
coordinates in   $ \gamma$ give rise to the stationary area condition for the
string.

The Einstein equations  still admit ordinary   black hole solutions
corresponding to zero string area.  The Euclidean space is regular at $r=2M$ and
the $t$-periodicity is the inverse Hawking temperature. However there exists a
non-trivial solution  to the string equations of motion in Euclidean space when
the string wraps around the Euclidean continuation of the horizon, a sphere at
$r=2M$. This solution has a curvature singularity at $r=2M$. Expressing the
curvature in the trace of Einstein equations as  the product of the
horizon times a two dimensional curvature and using the Gauss-Bonnet theorem
for disc topology  tell us that there is a conical singularity with
deficit angle $ 2 \pi \eta$ such that 
\begin{equation}  
\eta=4 \mu.
\end{equation}
This deficit angle is the sole effect of the string instanton.  It raised the
temperature from $\beta^{-1}_H$ to $\beta^{-1}=\beta_H^{-1} / (1-4 \mu)$.
 
I now evaluate the free energy of the black hole. The contribution of the
string term to the action  (\ref{action}) exactly cancels the
contribution of the Einstein term. The only
contributions comes from the boundary terms and one gets 
\begin{equation}
\label{free}
F(\beta,\mu) = \beta^{-1}   I_{saddle} = {M \over 2}= {\beta \over 16 \pi
(1-4\mu)}. 
\end{equation}
From Eq. (\ref{free}) and the thermodynamic relations $S = \beta^2   
(\partial F / \partial \beta)_\mu$, $ X_{4\mu}= (\partial F / \partial
4\mu)_\beta $, one recovers Eqs. (\ref{force}) and (\ref{areanew}) with
$\eta=4\mu$.

For a black hole of given mass $M$, the increase of the Hawking temperature due
to the conical singularity is accompanied by a decrease of the entropy in such a
way that the product 
\begin{equation}
TS = {M \over 2} 
\end{equation}
remains constant. This result, which is consistent with the fact mentioned
before that the classical differential mass formula (\ref{dmf}) defines the
entropy only up to a multiplicative constant, can be understood in simple
terms. Hawking quanta carry away a mass   proportional to the
temperature but the entropy contained in the radiation is proportional to the
number of emitted quanta. Hence   the entropy stored in a
black hole of mass $M$ must decrease as the   temperature increases. 

In the above analysis I   considered the entropy associated to an
instanton corresponding to a string wrapped once around the horizon. Clearly
such a configuration is thermodynamically meaningful only if it describes  a
metastable  state at the scale of the black hole lifetime.
Otherwise the instanton, as well as  instanton configurations with higher
winding number would contribute only an irrelevant exponentially small correction
to the leading entropy term $A/4$.  I shall assume temporarily that such a  
metastability does indeed occur and investigate its consequences at a quantum
level. This will allow me to exhibit a mechanism which would lead to a unitary
evolution of the collapsing black hole and which could well be realised by any
phenomenon implying  non local correlations on the
scale of the horizon.

\subsection{Towards unitarity.}

The  string instanton   at $r=2M$ in Euclidean space does not alter
the classical Lorentzian black hole background  which remains
regular on the horizon.   However dramatic   effects occur  at the
quantum level. To illustrate these let us consider the approximation
consisting of retaining only the $s$-wave component of a free scalar field
propagating on the Schwarzschild  geometry and   disregarding the residual
relativistic potential barrier. This amounts to take a
2-dimensional scalar field  propagating on the radial subspace of
the 4-geometry. The metric is
\begin{equation}
\label{metric}
ds^2= -( 1- {2M\over r}) du dv \qquad  u =t-r^*, v=t+r^* 
\end{equation}
where $r^*$ is the tortoise coordinate
\begin{equation}
\label{tortoise}
  dr^* = dr (1-2M/r)^{-1}.
\end {equation} 
One can    compute   the  expectation value of the
energy-momentum tensor of the scalar field using the trace anomaly \cite{dfu}
to get
\begin{equation}
\label{trace}
4\pi r^2 T_{uu} = {1\over 12 \pi} \left[-{M\over 2r^3}( 1- {2M\over r}) -
{M^2\over 4r^4}\right] + t_u(u).
\end {equation}
Here $t_u(u)$ is defined by   boundary conditions. For a radiation flux at a
temperature $T$, one has  
\begin{equation} 
\label{boundary}
\lim_{r\to \infty} 4\pi r^2 T_{uu}=t_u=(\pi/12)T^2.
\end{equation}
Inserting this value into Eq. (\ref{trace}) one sees that if $T=T_H$, $T_{uu}$
vanishes on the horizon as $(1-2M/r)^2$. This means that in the Kruskal inertial
frame defined by
\begin{eqnarray}
dU&=& \exp ({-u\over 4M}) du \\
dV&=&\exp ({v\over 4M}) dv,
\label{kruskal}
\end{eqnarray}
the energy momentum tensor is regular on the horizon. If    the
temperature is increased above $T_H$, one gets 
\begin{equation}
\label{singular} 
4\pi r^2 T _{UU}= {1\over 48 \pi}\left[({T\over T_H})^2 -1\right] {1\over U^2}.
\end {equation}
Thus, the string instanton induces, in an inertial frame,   a
positive energy singularity   on the horizon.

Let us suppose that the above analysis of an eternal black hole can be extended
to the radiation emitted by a black hole originating from a collapse. In this
case the past horizon is absent but the behaviour on the future horizon
  is unaltered. Hence, in absence of backreaction, the outgoing flux is still 
given by Eqs.(\ref{trace}), (\ref{boundary}) and (\ref {singular}) at
asymptotic Schwarzschild times. 

 To estimate the   backreaction, I shall compute the amount of
radiation energy stemming from the region just outside the star. In absence of
backreaction, a collapsing shell moves at asymptotic times along a trajectory
of constant $v=v_0$ in the coordinate system defined by Eqs.(\ref{metric}) and
(\ref{tortoise}) which remain valid outside the shell. The latter can be
taken as a mimic of the star surface. Outside the star, energy is conserved and
the energy flux measured at asymptotic times as $r \to \infty$  (or $v \to
\infty$) stems entirely from the negative energy flux $T_{vv}$ flowing towards
the horizon. No radiation flux $T_{uu}$ is emitted from the vicinity of
the star.  This is why the latter is not affected by the radiated quanta and
why no information about it is carried by the radiation. When the temperature is
increased, the vicinity of the star, contributes to the radiation an energy
flux which from Eqs.(\ref{trace}) and (\ref{boundary}) is
\begin{equation}
 4\pi r^2 T_{uu}= {\alpha \over M^2}, \qquad \alpha= M^2{\pi\over
12}(T^2-T_H^2) ={\mu (1-2\mu)\over 96\pi (1-4\mu)^2}.
\label{starradiation}
\end {equation}

It follows from Eq.(\ref{starradiation}) that, if $\alpha$ is of $O(1)$, the
radiation stemming from the region just ouside the star would tend to
infinity as $u\to\infty $. Hence one expects that quantum gravity must deeply
affect  the star and its   neighbourhood.   I shall label these as   the
quantum star. Even without the knowledge of the equation of state of the quantum
star one may use mass conservation to estimate its
average coordinate distance $y\equiv r(t) - 2 M(t)$to its Schwarzschild
radius.   In view of  Eq.(\ref{starradiation}), the mass M(t) of the quantum
star decreases in time as $dM/dt\simeq  -\alpha/M^2 $. As a first appromimation
I keep  $v=v_0 =$ constant and thus, from Eqs.(\ref{metric})and
(\ref{tortoise}), one gets when $y \ll {2M}$
\begin{equation} 
 \label{planck}
{dy\over dt} \simeq {-y\over 2M} + {\alpha \over M^2}.
\end {equation}

Equation (\ref{planck}) is similar to a result obtained  by
Itzhaki \cite{itzhaki}  in a different but related context. We see
  that in a time comparable to $4M \ln M$ after the onset of the Hawking
radiation which occurs when $y$ is still of order $M$, the quantum star reaches
a coordinate distance $y=O(1/M)$ of its Schwartzschild radius.
The proper distance $d$ to the horizon in the Schwarzschild time   is 
\begin{equation}
 d=\int_{2M}^{r}\  {1\over \sqrt{1 -{2M \over  x} }} d x   \simeq \sqrt{8My} =
O(1). \end{equation}
Thus the quantum star remains at a Planckian distance (and at a Planckian
temperature) of its Schwarzschild radius up to a time of order $M^3$ where it
has completely evaporated. After the   time $4M \ln M$, the nature of the Hawking
radiation changes: thermal quanta are now emitted from the burning   quantum
star  and    information about its structure gets transmitted to the
radiation. These considerations corroborate the results of a previous
analysis based on the assumption that quantum gravity should tame the
transplanckian frequencies of blueshifted Hawking quanta stemming from  the
neighbourhood of the horizon after such a time span \cite{eng}.

Thus the would-be horizon in absence of backreaction  may disappear and mass
conservation alone indicates that the incipient black hole may  evaporate
before a true horizon can form. Such a history of a collapsing star in a
topologically trivial background devoid of horizon and singularity would clearly
 be consistent with unitarity. The above analysis indicates that this is likely
to be the case if, as exemplified by the string instanton, non local
correlation's on the scale of the horizon induce at the quantum level a non
vanishing positive energy density $T_{uu}= O(1/M^2)$  when $R\to 2M$.

\section{The Complementarity Issue}

We just saw that unitarity could be maintained in black hole evaporation if 
the energy density $T_{uu}$ remains finite when $R \to 2M$. The
halting mechanism resulting from the non vanishing $T_{uu}$  would also
solve, as mentioned above, the well known problem posed
by the existence of transplanckian frequencies in an external observer frame.
However,  it   apparently contradicts the equivalence principle as the small
classical gravitational field at the horizon appears inconsistent with the huge
acceleration required to bring the collapse to a halt. 

 This clash could be resolved in quantum gravity where the metric field
$g_{\mu\nu}(x)$ is promoted to a   quantum operator. Indeed, to the extend that
a description in terms of matter evolving in a geometrical background 
 could be maintained, the metric $g_{\mu\nu}(x)$ need not be identified with 
the expectation value of the corresponding
Heisenberg  operator $\hat g_{\mu\nu}(x)$ in a   ``in''
quantum state $\ket {i} $.   Rather, the background geometry could
be determined, as anticipated by 't Hooft \cite{thooft2}, from both   the ``in''
state $\ket {i} $   and the ``out'' state. This would lead to different 
causal histories of the black hole as reconstructed by observers 
crossing the horizon at different times but would reduce, in accordance
with the equivalence principle, to the classical description of the
collapse for the proper history of the star as recorded by an observer
comoving with it \cite{eng}.

To understand this point in qualitative terms, consider a detector
sensitive only to cisplanckian effects. I call such a detector an
observer.  Let us first confine the motion of this observer within the
space-time outside the event horizon of a sufficiently massive
collapsing star. It  will necessarily encounter radiation. The 
radiation recorded by such   observers can be encoded in some
``out''-state . Thus  in the space-time available to ``external''
observers, there exist  outgoing states describing  a particular set of
detectable  quanta covering the whole history of the evaporating black
hole.  This information  about a particular decay mode can be added to
the characterisation of the system by the Schr\"odinger state of the
star  before collapse, or equivalently by the corresponding Heisenberg
state $\ket {i}$. More precisely we could specify that the system is
likely to be  found at sufficiently late times in a state
characterised by some typical distribution of Hawking quanta. It may
seem at first sight that this added information about the future
detection of Hawking radiation is irrelevant for the   analysis of the
energy momentum tensor and of the metric at intermediate times. This
could be an incorrect conclusion for reasons I shall now explain.
 
The expectation value $ \bra i \hat A(t) \ket i$  of a Heisenberg
operator $\hat A(t)$ in the normalised  quantum state $\ket i $ is often
expressed as
\begin{equation}
     \bra i \hat A(t) \ket i =\sum_\alpha
P_\alpha\,A_\alpha \qquad   P_\alpha =\vert\langle \alpha  \ket i\vert^2
\end{equation}
where the eigenvectors $\ket \alpha$ relative to eigenvalues
$A_\alpha$ form a complete set of orthonormal states. One   interprets
then  $  \bra i \hat A(t) \ket i$  as the average over the probability 
distribution $P_\alpha$ of finding the value $A_\alpha$ if exact
measurements of a   complete set of commuting observables containing
$\hat A(t)$  are performed at time $t$ on a quantum system
``pre-selected'' to be in the initial Shroedinger state $\ket{t_1,i} =
U(t_1,t_0)\ket i$ at time $t_1$. $U(t_1,t_0)$ is the evolution operator
to the time  $t_1$ from the time $t_0$ where the Shroedinger state is
identified with the Heisenberg one.
 
More information can   be gained if the system is also
``post-selected'' to be found at a later time $t_2$ in a given
Schr\"odinger state $\ket{t_2,f}$ \cite{aharonov}.  One may then express
expectation values  $\bra i \hat A(t) \ket i$ as an average of   weak
values defined for $ t_1<t<t_2$ by
\begin{equation}
\label{weak}
 A^{weak}_f  \equiv {\bra f \hat
A(t) \ket i \over \bra f i\rangle}
\end{equation} 
where $\ket f = U(t_0,t_2)
\ket{t_2,f}$.  
One gets  
\begin{equation}
\label{average}
  \bra i \hat A(t) \ket i = \sum_f
P_f\,A^{weak}_f \qquad   P_f =\vert\langle f \ket i\vert^2. 
\end {equation}
Eq.(\ref{average}) suggests that weak values represent   measurable quantities
for a pre- and post-selected system. This is indeed the case if a measurement
of $\hat A(t)$ is performed on the system with sufficient quantum
uncertainty to avoid disrupting the evolution of the system. Such
``weak'' measurements yield not only the real part of  $A^{weak}_f$ but
also its imaginary part and reconstruct in this way the available
history between $t_1$ and $t_2$ for a system  pre-selected at $t_1$ and
post-selected at $t_2$ \cite{aharonov}. In the limit $\hbar \to 0$,
the weak value is purely real.
 
Generally, the information gained by post-selection and weak
values is relevant only if one post-selects a state   describing   rare
events for the pre-selected state considered, that is if $P_f$ is
located in the tail of the distribution probability. However, the
situation is different when one considers the Hawking emission process in
the classical background of a collapsing star. There, in absence of
backreaction, the energy-momentum tensor of the radiation can be
computed exactly in some simplified  models. It then appears
that post-selected states defined on a space-like surface $\cal \sigma$
arbitrarily close to the union of the event horizon   and of the
future light-like infinity  may yield weak values of the
energy-momentum tensor operator $\hat T_{\mu\nu}(x)$ very different 
from its average value. While the latter remains  smooth on the scale of
the Schwartzschild   radius, the former may  exhibit in that region
oscillations of unbounded amplitudes \cite{massar1,massar2}. These features,
which are a  consequence of the unbounded blueshift experienced in the vicinity
of the horizon by the vacuum fluctuations generating the Hawking quanta, persist
in   generic post-selected states detectable by external observers.  The
energy content of these  fluctuations show up in the weak values of
$\hat T_{\mu\nu}(x)$ but are averaged out in expectation values. However
observers who do cross the horizon detect different post-selected
states. These yield weak values of  $\hat T_{\mu\nu}(x)$  which 
are  smooth as the observer approaches the horizon. In particular free falling
observers cannot detect Hawking quanta and their weak values coincide with the
expectation values \cite{eng}. This suggests that, when backreaction is taken
onto account, there is no reason to identify for all observers, the relevant
metric background to the expectation values of the metric operators $\hat
g_{\mu\nu}(x)$.  

Transplanckian effets and unitarity
considerations pose problems only to the external observer who can ultimately
detect Hawking quanta   encoded in some final state $\ket f$. Both problems
would be cured by the halting mechanism as the latter would tame the
transplanckian fluctuations of the weak value in $\hat T_{\mu\nu}(x)$. This
raises the possibility that the halting mechanism is a property of the weak
value of the metric relative to the final state  $\ket f$. In other words,
halting would occur only in the
description of the {\em reconstructed} history encoded in the background defined
by the weak value of the metric
\begin {equation}
\label{weakg}
 g^{rec}_{\mu\nu}(x)={\langle f \vert \hat g_{\mu\nu}(x)\vert i \rangle
\over \langle f \vert i\rangle}.
\end {equation}
The inertial observer, which cannot detect Hawking quanta, would observe the 
{\em proper} history encoded in the expectation value
\begin {equation}
\label{proper}
 g^{proper}_{\mu\nu}(x)=\langle i \vert \hat g_{\mu\nu}(x)\vert i\rangle.
\end {equation}

The above discussion indicates that  black hole complementarity \cite{suss} can
fit into the framework of conventional quantum physics. It could reconcile  
non local correlations on the horizon scale required for the unitarity of
the reconstructed history with the equivalence principle as applied to the
proper history. While the  correlations  necessary for ensuring unitarity appear
hardly available in the context of conventional local field theory, there
are some  indications that the developments of string theories  point
towards a unitary description of black hole decay. But a full non perturbative
formulation is clearly required to settle the   issues.

\subsection*{Acknowledgements}

I am grateful to R. Argurio and L. Houart for valuable discussions.


\begin{thebibliography}{9}
  
 
\bibitem{vafa} A. Strominger and C. Vafa, Phys. Lett. B379 (1996) 99,
hep-th/9601029.

\bibitem{callan} C. G. Callan and J. M. Maldacena, Nucl. Phys. B472 (1996) 591,
hep-th/9602043.

\bibitem{bek} J. D. Bekenstein, Phys. Rev. D7 (1973) 2333.

\bibitem{haw} S. W. Hawking, Commun. Math. Phys. 43 (1975) 199.  

\bibitem{haw2} S. W. Hawking, Phys. Rev. D14 (1976) 2460.

\bibitem{remnant}  Y. Aharonov, A. Casher and S. Nussinov, Phys. Lett. 
B191 (1987) 51.

\bibitem{thooft} G.'t Hooft,  Nucl. Phys. B256 (1985) 727. 

\bibitem{suss} L. Susskind, L. Thorlacius and J. Uglum, Phys. Rev. 
D48 (1993) 3743, hep-th/9306069.  

\bibitem{bch} J. M. Bardeen, B. Carter and S. W. Hawking, Comm. Math. Phys. 
31 (1973) 161.

\bibitem{ce} A. Casher and F. Englert, Class. Quantum Grav. 9 (1992) 2231. 

\bibitem{gh1} G. W. Gibbons and S. W. Hawking, Phys. Rev. D15 (1977) 2752.

\bibitem{ehw1} F. Englert, L. Houart and P. Windey, Phys. Lett. B372 (1996) 29,
 hep-th/9503202.

\bibitem{ehw2} F. Englert, L. Houart and P. Windey, Nucl. Phys. B458 (1996) 231,
 hep-th/9507061.

\bibitem{dfu}  P. C. W.  Davies, S. A. Fulling and W. G. Unruh, Phys.
Rev. D13(1976)2720.

\bibitem{itzhaki} N. Itzhaki, ``Is the Black Hole Complementarity Principle
Really Necessary'', hep-th/9607028.
 

\bibitem{eng} F. Englert, ``Operator Weak Values and Black Hole
 Complementarity'',  Proc. of the Oskar Klein Centenary
 Symposium, edited by U. Lundstroem (World Scientific, Singapore, 1994),
gr-qc/9502039.


\bibitem{thooft2} G.'t Hooft, ``Unitarity of the Black Hole Scattering Matrix''
 Proc. of the ``International Conference on Fundamental Aspects of
Quantum Theory'' (1992), in Honour of Y. Aharonov's 60th birthday.
\hfill \break\  C. R. Stephens, G.'t Hooft and B. F. Whiting,  Class. Quantum
Grav. 11 (1994) 621, gr-qc/9310006.


\bibitem{aharonov} Y. Aharonov, D. Albert, A. Casher and L. Vaidman, Phys.
Lett. A124 (1987) 199. \hfill \break Y. Aharonov and L. Vaidman,
Phys. Rev. A41 (1990) 11. \hfill \break Y. Aharonov, J. Anandan,
S. Popescu and L. Vaidman, Phys. Rev. Lett.  64 (1990) 2965.

\bibitem{massar1}  S. Massar and R. Parentani, ``From vacuum Fluctuations to
Black Hole Radiation'', gr-qc/9404057. \hfill \break\  Phys. Rev.
D54 (1996) 7426, gr-qc/9502024.

\bibitem{massar2} F. Englert, S. Massar and R. Parentani, Class. Quantum
Grav. 11 (1994) 2919, gr-qc/9404026.






 

\end{thebibliography}
\end{document}